\documentclass[11pt, a4paper]{amsart}

\usepackage{amsmath}
\usepackage{amsthm}
\usepackage{amssymb}
\usepackage{enumerate}
\usepackage{hyperref}
\usepackage{amsaddr}

\begin{document}
	
\title{The ``paradox'' of computability and a recursive relative version of the Busy Beaver function}

\author[fsa]{Felipe S. Abrah\~{a}o}
\address[fsa]{National Laboratory for Scientific Computing (LNCC), Brazil}

\thanks{This research was supported by the National Council for Scientific and Technological Development (CNPq), Brazil, as a PhD Fellowship at the Federal University of Rio de Janeiro (UFRJ), Rio de Janeiro, Brazil.}
\thanks{This is a preprint of a book chapter published in \textit{Information and Complexity}, Mark Burgin and Cristian S. Calude (Editors), World Scientific Publishing, 2016, ISBN 978-981-3109-02-5, available at \url{http://www.worldscientific.com/worldscibooks/10.1142/10017}.}

\begin{abstract}
	In this article, we will show that uncomputability is a relative property not only of\textbf{\textit{ }}oracle Turing machines, but also of\textbf{\textit{ }}subrecursive classes. We will define the concept of a Turing submachine, and a recursive relative version for the Busy Beaver function which we will call Busy Beaver Plus function. Therefore, we will prove that the computable Busy Beaver Plus function defined on\textbf{\textit{ }}any Turing submachine is not computable by\textbf{\textit{ }}any program running on this submachine. We will thereby demonstrate the existence of a ``paradox'' of computability \textit{a la} Skolem: a function is computable when ``seen from the outside'' the subsystem, but uncomputable when ``seen from within'' the same subsystem. Finally, we will raise the possibility of defining universal submachines, and a hierarchy of negative Turing degrees.
\end{abstract} 

\maketitle
	
\section{Introduction}\label{sec1.1}

	In first place, we must briefly introduce the ideas behind the definitions, concepts and theorems exposed in the present article. It is true that, following an interdisciplinary course of study, this work focuses on manifold inspirations coming from several different fields of knowledge. However, for present purposes it is essential to mention the two foremost: Skolem's ``paradox'' and metabiology.
	
	Skolem's ``paradox'' derives from Cantor's famous theorem on the uncountablity of infinite sets, for instance of real numbers or of the set of all subsets of natural numbers, and also from L\"{o}wenheim-Skolem's theorem on the size of models in satisfiable theories. Briefly, the ``paradox'' is: there is a countably infinite model for a theory (e.g. ZFC, if we accept it as consistent) that proves there are uncountably infinite sets. Therefore, when we look at the set of elements in the theory's model which correspond homomorphically to the\textbf{\textit{ }}elements in the set that the theory demonstrates contain an uncountable amount of elements, may however contain a countable amount of elements. Does this contradiction represent, in fact, a paradox?${}^{ }$
	
	That question was answered by Skolem in 1922, being that the property of countability of real numbers depends on the existence of a function within the model that makes this bijective enumeration. This function cannot exist within any\textbf{\textit{ }}ZFC model - if that is the chosen axiomatic system - because, if it existed, it would render countable the set of all real numbers. However, it may exist\textbf{\textit{ }}when seen ``from the outside'', when this function never belongs to the model. In other words, the function that ``bijects'' the natural onto the real numbers never belongs to any model of any Set Theory\footnote{ Of course, provided it is strong enough and $  $satisfiable$  $.} - but exists nevertheless. Thus, it is understood that this does not constitute a true paradox. It forms what one may call a pseudoparadox: when seen ``from the outside'', an object has a certain property, but when seen seem ``from within'' it has the opposite property. For this reason, it makes sense to call the Skolem's ``paradox'' a pseudoparadox of countability. Therefore, one may ask the question: just as there is a pseudoparadox of countability, could there be a pseudoparadox of computability? We will address this subject in the present paper.
	
	Metabiology is a field of theoretical computer science with a transdisciplinary ``heart'' that studies general principles of biological relations at a meta-level focusing on the open-ended evolution of systems and is inspired by both the theories of biological evolution and by algorithmic information theory (AIT). While the already developed and well-established fields of population genetics and evolutionary computations are driven towards simulations of evolutionary populations and statistical properties of those populations, metabiology is driven towards achieving theorems. It uses all available tools from theory of computation, algorithmic information and metamathematics to build and study abstract models -- applicable or not.
	
	Unlike the first models made by Chaitin, if we want a ``nature'' without access to oracles - at least, without access to real\textbf{\textit{ }}oracles - it needs to be a system that can be completely simulated on a universal Turing machine, or on a sufficiently powerful computer. This would also allow us to do experimental computer simulations on the evolution of digital organisms in the future.
	
	In attempting to create a metabiological nature which is computable, but that ``behaves in the same way'' as an oracular one, i.e., in the same way as an uncomputable nature - in other words, as a hypercomputer - we face a series of difficulties. The first, highlighted in the literature, is the absence of a program that will solve the \textit{halting problem}. This is required to determine whether or not an algorithmic mutation will result in\textbf{\textit{ }}a new organism/program (a mutation's output given by a prior organism as input) and whether or not the organism/program has a higher fitness (its output) than the previous one. A computable nature would necessarily need to be capable of running a function capable of accomplishing this task, which we know is impossible for any arbitrary program. Note also that solving the halting problem, computing the bits of a Chaitin number $\mathrm{\Omega }$\textbf{\textit{, }}and computing the Busy Beaver function, are equivalently uncomputable problems of Turing degree $\boldsymbol{0'}$. How would evolution, then, occur within a computable simulation of ``Nature''?
	
	However, this paper does not intend to present mathematical exercises in metabiology, but to focus on presenting the elements which allow proving a computable metabiological model: ``\textbf{sub-uncomputability''}. The present paper comes from a solution we gave to the question to the above, as shown in [4]. We intend to demonstrate several fruitful properties and raise issues for future study within the mathematics of theoretical computer science.
	
	Basically, we will call a \textbf{Turing submachine} any Turing machine that always gives an output for any input, i.e. always halts. The prefix ``sub'' parallels the term subrecursion. A subrecursive class is one defined by a proper subset of the set of all problems with Turing degree $\boldsymbol{0}$\textbf{.} Therefore, a \textbf{subcomputable\textit{ }}class of problems will be subrecursive, because it will never contain all recursive/computable problems. The term subrecursion is also used to characterize subrecursive hierarchies, as in Kleene and Grzegorczyk, covering all primitive recursive functions. But for us, the prefix refers more specifically to the concept of subrecursive class.
	
	Note that Turing submachine is just another terminology for \textbf{total Turing machines}. However, despite the fact that they are just different names for the same object and can be used interchangeably, the expression ``total Turing machine'' might not immediately capture its relevant properties related to the present paper. 
	
	Every total computable function (or total Turing machine) defines a subrecursive class which is a proper subclass of other subrecursive classes (and of the class of all recursive functions). For example, the total Turing machine will be a subsystem of another total machine which is capable of computing functions that are relatively uncomputable by the former. This very idea of being part of another non-reducibly more powerful machine -- that comes from sub-uncomputability, as we will show -- is the core notion of the expression ``submachine'', conveying and bearing the ideas of hierarchies of subrecursive classes together with the powerful concept of Turing machines. Thus, the terminology Turing submachine emphasizes this property of total Turing machines being always able to be part of another proper and bigger machine. For more of this discussion, see item 4.
	
	The central theme is building, or rather proving, a system (a Turing machine) that can ``behave'' in relation to a subsystem (its Turing submachine) in the same way as a hypercomputer (an oracle Turing machine) would behave in relation to a subsystem (in particular, a universal Turing machine). In fact, we will not emulate all - which might be impossible - the properties of a hypercomputer in relation to a computer, but focus on defining a function ${{BB}^+}_{{P'\textrm{´}}_T}(N)$ analogous to a Busy Beaver\textbf{\textit{ }}function (in Chaitin's work, function $BB'(N)$), so that this function will behave in relation to the Turing submachine$\ $in the ``same'' way as the original Busy Beaver behaves in relation to a universal Turing machine. In other words,  ${{BB}^+}_{{P'\textrm{´}}_T}(N)$ must be relatively uncomputable by any \textbf{subprogram} (a program running on a Turing submachine), the same way the original Busy Beaver is uncomputable by any program running on a universal Turing machine. This phenomenon will be called \textbf{recursive relative uncomputability, }or\textbf{ sub-uncomputability.}
	
	\section[Language L]{Language $L$}\label{sec1.2}
	The first important definition that needs to be established is the very programming, or universal machine, language with which we will work. It is important to us that the submachine $U_{{P'\textrm{´}}_T}$ must be programmable. This language can be used on any usual computer, that is, its properties and rules of well-formation are programmable. Ultimately, this will lead us to the conclusion that the phenomenon of subcomputation may occur ``within'' computers as we already know them, which are universal Turing machines with limited resources.
	
	Why concatenations? They provide a direct way of symbolizing a program taking any given bit string as input - for example, a program $p$ that is, actually, program $p'\textrm{´}\ $taking program $p''\textrm{´}\textrm{´}$ as input - which makes this program act as a function. Note that this type of program is already used to demonstrate the \textit{halting problem}, or demonstrate that the Busy Beaver function is uncomputable. But the form it may assume is completely arbitrary, as a universal Turing machine, in any case, will run it. Therefore, it is no wonder we need this condition -- this functionalizing special concatenation - in our language. As we are trying to build a computer that can emulate uncomputability, it is necessary that we can ``teach'' a machine to perform and recognize these ``concatenations'' within the language it is working in.
	
	Many of the properties of the language $L$, below, are not required for this study; however, they were required to demonstrate the evolution of metabiological subprograms \textit{a la} Chaitin.
	
	To differentiate from the optimal functionalizing concatenation, which is joining strings in the most compressed way possible, provided it remains well-formulated, this special functionalizing concatenation will be denoted as ``$\circ $'', while the optimal functionalizing concatenation will be symbolized as ``$*$''.
	
	\subsection{Definition}\index{Language $L$!Definition}
	We say a universal programming language $l$, defined on a universal Turing machine $U$, is \textbf{recursively functionalizable} if there is a program that, given any bit strings\textit{ }$P$ and $w\ $as inputs\textit{,} will return a bit string belonging to $l$ which will be denoted as $P\circ w$, whereby $U\left(P\circ w\right)\ $equals ``the result of the computation (on $U$) of program $P$ when $w$ is given as input''. In addition, there must be a program that will determine whether or not a bit string is in form $P\circ w$ for any $P$ and $w$, and is capable of returning $P$ and $w$  separately. Analogously, the latter must be true for the successive concatenation $P\circ w_1\circ \dots \circ w_k$, with program $P$ receiving $w_1$,{\dots}$w_k$ as inputs. 
	
	Now, the general definition of language $L$ can be defined as: let $U$ be a universal Turing machine running language $L$, a universal language that is \textbf{binary, self-delimiting, recursive,} and \textbf{recursively functionalizable  }and that there are constants $\in $, $C$ e $C'$, for every $P$, $w_1$,{\dots}$w_k$, where:

	\[\left|w_i\right|\mathrm{<}\left|P\mathrm{\circ }w_{\mathrm{1}}\mathrm{\circ }\mathrm{\dots }\mathrm{\circ }w_k\right|,for\ i\mathrm{=1,2,\dots }or\mathrm{\ }k\]

	\noindent and
	
	\noindent 
	\[\left|P\mathrm{\circ }w_{\mathrm{1}}\mathrm{\circ }\mathrm{\dots }\mathrm{\circ }w_k\right|\mathrm{\le }C\mathrm{\times }k\mathrm{+}\left|P\right|\mathrm{+}\left|w_{\mathrm{1}}\right|\mathrm{+|}w_{\mathrm{2}}\mathrm{|+\dots +|}w_k\mathrm{|}\]

	\noindent and
	
	\noindent 
	\[H\left(N\right)\mathrm{\le }C\mathrm{'+}{{\mathrm{log}}_{\mathrm{2}} N\ }\mathrm{+}\left(\mathrm{1+}\epsilon \right){{\mathrm{log}}_{\mathrm{2}} \left({{\mathrm{log}}_{\mathrm{2}} N\ }\right)\ }\] \\ \\
	
\section{Definitions}

\begin{enumerate}[(a)]
	\item \textbf{ }$W$ is the set of all finite bit strings, where the computable enumeration of these bit\textit{ }strings has the form $l_1$, $l_2$, $l_3$, ..., $l_k$, ...\textit{ }

	\noindent 
	
	\noindent For practical purposes, a language may be adopted where $l_1=0$.
	
	\noindent

	\item  Let\textbf{\textit{ }}$w\in W$.

	\noindent $|w|$ denotes the size or number of bits contained in \textit{w.}  
	
	\item  Let $N$ simply symbolize the corresponding program in language $L$ for the natural number $N$. For example, $P\circ N$ denotes program $P\circ w$ where $w$ is the natural number $N$ in the language $L$.

	\noindent

	\item  If function $f$ is computable by program $P$, then $f$ may also be called function $P$.
\end{enumerate}

\section{Turing Submachines}
A key concept in the present article is the idea that a subsystem can do almost anything its system can, however, with resources limited by the very system. We follow the conventional understanding in which a computation that is a part of another computation may be called a subcomputation, and a machine that is a part of another machine may be called a submachine. In our case, a system can be taken as a Turing machine, and a subsystem can be taken as a Turing submachine. For example, a Turing submachine can be a program or subroutine that the ``bigger'' Turing machine runs, always generating an output, while performing various other tasks. Note that it is true (a theorem) that for every total Turing machine there is another Turing machine that completely emulates and contains the former total Turing machine, in a manner that the computations of the latter contains the computations of the former.

In fact, we are using a stronger notion of subsystem based upon this conventional notion: a subsystem must be only able do what the system knows, determines and delimits. This way, submachines will only be those machines for which there is another ``bigger'' machine that can decide what is the output of the former and whether there is an output at all. Note that every machine that falls under this definition always defines an equivalent \textbf{total Turing machine} (with a signed output corresponding to the case where the former does not halt); and every total Turing machine falls under this definition. 

We will use another concept of vital importance: \textbf{ computation time.} Similarly to time complexity, we will call $T$ a program that calculates how many steps or basic operations $U$ performs when running program $p$. Thus, if $U(p)$ does not halt, then $U(T*p)$ will not halt either, and vice versa.

Let $P_f$ be\textit{ }a program running on $U$ defined in$\ $language$\ L$, computing a total function (a function defined for all possible input values) $f$ such that $f:L\longrightarrow X\subseteq W$. The language $W$ does not need necessarily to be self-delimiting, and may be comprised of all bit strings of finite size, as long as they may be recursively enumerated in order, as $l_1$, $l_2$, $l_3$,\textit{... }For practical reasons, we will choose an enumeration where $l_1=0$.

A \textbf{``Turing submachine'' or total Turing machine} ${U}/{f}$ is defined\textit{ }as a Turing machine in which, for every bit string $w$ in the language of $U$, ${U}/{f}\left(w\right)=U\left(P_f\circ w\right)$. 

This definition is quite general and transforms any total computable function into a Turing submachine. In fact, as said in the introduction, Turing submachines are just another name for total Turing machines. Anyway, Turing submachines can always be subsystems of either abstract universal Turing machines or of powerful (big) enough everyday computers (which are also some sort of total Turing machine, i.e. a universal Turing machine with limited resources).

Note that the class of all submachines is infinite, but not recursive.

When we talk of \textbf{subprograms }we refer to programs run on a Turing submachine. Herein, only submachines of a particular subclass will be dealt with: submachines defined by a computation time limited by a computable function\footnote{ Or time-bounded Turing machines}. In fact, both these and the more generic submachines defined above are equivalent in computational power. To demonstrate this, just note that if a program computes a total function, then there is a program that can compute the computation time of this first program. Therefore, for every computable and total\textbf{\textit{ }}function, there is a submachine with limited computation time capable of computing this function -- and, possibly, other functions as well. The reverse follows from the definition of submachine.

Let $P_T$ be\textit{ }an arbitrary program that calculates a computation time for a given program $w$. That is, let\textit{ }$P_T$ be\textit{ }an arbitrary total\textbf{\textit{ }}computable function. Thus, there is a \textbf{Turing submachine} ${\boldsymbol{U}}_{{\boldsymbol{P}}_{\boldsymbol{T}}}$ \textbf{defined by the computation-time function }${\boldsymbol{P}}_{\boldsymbol{T}}$.

We then define submachine ${U}/{{P_{SM}\circ P}_T}$ (which will be a program running on $U$ that computes a total\textbf{\textit{ }}function), where $P_{SM}$ is a program that receives $P_T$ and $w$ as inputs, runs $U(P_T\circ w)$ and returns:\\

\begin{enumerate}[(i)]
	\item  $l_1$, if $U(w)$ does not halt within computation time $\le U(P_T\circ w)$;
	
	\item  $l_{k\mathrm{+1}}$, if $U\mathrm{(}w\mathrm{)}$ halts within computation time $\mathrm{\le }U\mathrm{(}P_T\mathrm{\circ }w\mathrm{)}$ e $U\mathrm{(}w\mathrm{)=}l_k$;\\
\end{enumerate}

This program defines a Turing submachine that returns a known symbol (in this case, zero) when program $w$ does not halt in time $\le U\left(P_T\circ w\right)\ $or returns the same output\textit{ }(except for a trivial bijection) as $U\left(w\right)$ when the latter halts in time $\le U(P_T\circ w)$.

To be a Turing submachine, ${U}/{{P_{SM}\circ P}_T}$ must be defined for all inputs. This occurs because  $P_T$ is total by definition. In addition, as computation time $P_T$ becomes more increasing, the more submachine\textit{ }${U}/{{P_{SM}\circ P}_T}$ approaches the universality of $U$\textit{.}

Therefore, we will denote only as $U_{P_T}$ a Turing submachine\textit{ }${U}/{{P_{SM}\circ P}_T},$\textit{ }so that:

\[\mathrm{\forall }w\in L\ \mathrm{(\ }U_{P_T}\left(w\right)\mathrm{=}{U}/{{P_{SM}\mathrm{\circ }P}_T}\left(w\right)\mathrm{=}U\left(P_{SM}\mathrm{\circ }P_T\mathrm{\circ }w\right)\mathrm{)}\]

\section[Function Busy Beaver Plus]{Function ${{BB}^+}_{{P'\textrm{´}}_T}(N)$}
Let ${P'\textrm{´}}_T\mathrm{\ }$be a total function. Let us define function ${{BB}^+}_{{P'\textrm{´}}_T}(N)$, which we will call Busy Beaver Plus, through the following recursive procedure:\\

\begin{enumerate}[(i)]
	
	\item  Generate a list of all outputs of $U_{{P'\textrm{´}}_T}\left(w\right)$ such that $\left|w\right|\le N$;
	
	\item  Take the largest number on that list;
	
	\item  Add 1;
	
	\item  Return that value.\\
	
\end{enumerate}

\noindent 

The name of this function refers to the Busy Beaver $BB\left(N\right)$ function and, consequently, it is no coincidence that the two have almost the same definition. If step (iii) is removed, it becomes exactly the Busy Beaver function for\textbf{\textit{ }}Turing submachines, here denoted as ${BB}_{{P'\textrm{´}}_T}(N)$. Thus:

\[{BB}^{\mathrm{+}}\left(N\right)\mathrm{=}BB\left(N\right)\mathrm{+1}\]

\noindent and

\[{{BB}^{\mathrm{+}}}_{{P'\mathrm{\textrm{´}}}_T}\left(N\right)\mathrm{=}{BB}_{{P'\mathrm{\textrm{´}}}_T}\left(N\right)\mathrm{+1}\]

But why use function ${BB}^+$ instead of $BB$? This might be, one supposes, the reader's first and immediate question. As we are dealing with Turing submachines and ${P'\textrm{´}}_T$ is arbitrary, it is possible there is a program on $U_{{P'\textrm{´}}_T}$ with size $\le N$ such that computes the highest value returned by any other program on $U_{{P'\textrm{´}}_T}$ with size $\le N$. When dealing with a universal Turing machine $U$, this cannot occur -- except for a constant. However, with submachines, it can. Thus, function ${{BB}^+}_{{P'\textrm{´}}_T}$ is triggered to assure it, in itself, is not relatively computable - or compressible -- by any program on $U_{{P'\textrm{´}}_T}$, although it can be by a program on$\ U$. Since ${P'\textrm{´}}_T$ is a program that computes a total\textbf{\textit{ }}function, then ${{BB}^{\mathrm{+}}}_{{P'\mathrm{\textrm{´}}}_T}\left(N\right)$ is computable.

The Busy Beaver contains the idea of the greatest output of any $\le N$ sized program; so the Busy Beaver Plus function contains the idea of increasing, at least by 1, any $\le N$ sized program. Respectively, the first gives us maximization, and the second, an ``almost'' minimal increment.

Following this line of thought, to symbolize this new function, the image may be evoked of the proverbial man sitting on a hungry donkey and driving the animal by a carrot hanging from a fishing rod. As the carrot looms in front of the donkey's face, the hungry donkey is driven to walk forward to reach the carrot, which is never reached. Not because it is an infinite distance away, but because with every step it takes the carrot moves forward along with it. The carrot is always ``one step'' ahead of the donkey. No matter how dutifully the donkey walks toward the carrot, it will always remain at the same distance, just beyond reach, unattainable. No matter how rapidly increasing is the function ${P'\textrm{´}}_T$, the program on $U$ that computes ${{BB}^+}_{{P'\textrm{´}}_T}\left(N\right)$ simply bases itself on the $U_{{P'\textrm{´}}_T}$ outputs to overcome them by a minimum. No matter how powerful $U_{{P'\textrm{´}}_T}$ may be, ${{BB}^+}_{{P'\textrm{´}}_T}\left(N\right)$ will always be ``one step'' ahead of the best that any subprogram (i.e., any program $U_{{P'\textrm{´}}_T}$) can do.

It is worthy of note that, analogously to the Busy Beaver, the ${{BB}^+}_{{P'\textrm{´}}_T}\left(N\right)$ may be used to measure the ``creativity'' or \textit{``sub-algorithmic complexity''} of the subprograms in relation to Turing submachine $U_{{P'\textrm{´}}_T}$. Why? By its very definition, if a subprogram generates an output $\ge {{BB}^+}_{{P'\textrm{´}}_T}\left(N\right)$, it must necessarily be of size $>N$. It needs to have over $N$ bits of relatively incompressible information, i.e. over $N$ bits of relative creativity. 

Of course, one may always build a program that will compute function ${{BB}^+}_{{P'\textrm{´}}_T}\left(N\right)$, if  function ${P'\textrm{´}}_T$ is computable. This would allow a far smaller program than $N$ there to exist -- e.g., of size $\le C+{{log}_2 N\ }+(1+\epsilon ){{log}_2 ({{log}_2 N\ })\ }$ -- that will compute ${{BB}^+}_{{P'\textrm{´}}_T}\left(N\right)$. But that does not constitute a contradiction, because this program can never be a subprogram of $U_{{P'\textrm{´}}_T}$, in other words, it can never be a program that runs on the computation time determined by ${P'\textrm{´}}_T$. If it was, it would enter into direct contradiction with the definition of ${{BB}^+}_{{P'\textrm{´}}_T}$: the program ${P'\textrm{´}}_T$ will become undefined for an input, which by assumption is false. Then, as we will show in item 6, we will immediately get the ``paradox'' of computability.

\section{Sub-uncomputability: Recursive Relative Uncomputability}
Now we will prove the crucial, yet simple, result that governs this paper. 

Let ${P'\textrm{´}}_T$ be a total function and $U_{{P'}_T}$ a Turing submachine. Then, we can prove that function ${{BB}^+}_{{P'}_T}\left(N\right)$ is \textbf{relatively uncomputable} by any program on $U_{{P'\textrm{´}}_T}$. Or: there is no subprogram that, for every input $N$, returns an  output equal to ${{BB}^+}_{{P'\textrm{´}}_T}\left(N\right)$. Actually, ${{BB}^+}_{{P'\textrm{´}}_T}\left(N\right)$ eventually dominates any program on $U_{{P'\textrm{´}}_T}$. 

A more intuitive way to understand what is going on is to look for a program and concatenate its input, such as $U_{{P'\textrm{´}}_T}\left(P*N\right)\ $for instance. Where ``$*$'' denotes the optimal functionalizing concatenation, and not necessarily the ``concatenation'' ``$\circ $'' defined in item 2. In fact, this applies to any way to compress the information of $P$ and $N$ in an arbitrary subprogram. Therefore, it may not be in the form $P\circ N$.

We avail ourselves of the same idea used in the demonstration of Chaitin's incompleteness theorem. Now, however, to demonstrate an uncomputability relative to the submachine $U_{{P'\textrm{´}}_T}$.

When $N$ is given as input to any program $P$, it comes in its compressed form with size $\cong H\left(N\right)$ -- in fact, we use the property 
\[\left|P\mathrm{\circ }N\right|\mathrm{\le }C\mathrm{+|}P\mathrm{|+}C\mathrm{'+}{{log}_{\mathrm{2}} N\ }\mathrm{+(1+}\epsilon \mathrm{)}{{log}_{\mathrm{2}} \mathrm{(}{{log}_{\mathrm{2}} N\ }\mathrm{)}\ }\]

 \noindent whereby $\left|P\circ N\right|\cong C+H\left(N\right).$ But, as already known by the AIT, for any constant $\mathrm{C}$\textbf{\textit{ }}there is a big enough $N_0\ $such that\textbf{\textit{ }}$C+H\left(N_0\right)<N_0$. Therefore, according to\textbf{\textit{ }}the definition of ${{BB}^+}_{{P'\textrm{´}}_T}$, the output of $P\circ N_0$ when run on submachine $U_{{P'\textrm{´}}_T}$, will be taken into account when one calculates ${{BB}^+}_{{P'\textrm{´}}_T}\left(N_0\mathrm{\ }\right)$. Thus, necessarily,

\[{{BB}^{\mathrm{+}}}_{{P'\mathrm{\textrm{´}}}_T}\left(N_0\mathrm{\ }\right)\mathrm{\ge }U_{{P'\mathrm{\textrm{´}}}_T}\left(P\mathrm{\circ }N_0\right)\mathrm{+1>}U_{{P'\mathrm{\textrm{´}}}_T}\left(P\mathrm{\circ }N_0\right)\]

\noindent Which will lead to contradiction, if $P$ computes ${{BB}^+}_{{P'\textrm{´}}_T}$ when running on submachine $U_{{P'\textrm{´}}_T}$. The same holds for ``$*$''.

Also, following the same argument, it can be shown promptly that ${{BB}^+}_{{P'\textrm{´}}_T}\left(N\right)$ is a \textbf{relatively incompressible, or sub-incompressible, }function by any subprogram smaller than or equal to $N$. That is, no program of size $\le N$ running on\textbf{\textit{ }}${\ P'\textrm{´}}_T$ will result in an output larger than or equal to ${{BB}^+}_{{P'\textrm{´}}_T}\left(N\right)$.

\section{Conclusion and Final Comments}
First, a self-delimiting universal language $L$ was defined for a universal Turing machine $U$. Then, we defined the Turing submachines (or total Turing machines) ${U_{P{'_T}}}$. It has been further demonstrated that the phenomenon of ``sub-uncomputability'' is ubiquitous: for every Turing submachine ${U_{P{{\text{'}}_T}}}$, if  is a program that computes a total function, then the computable function $B{B^ + }_{P{'_T}}\left( N \right)$ is relatively uncomputable by any program running on ${U_{P{{\text{'}}_T}}}$ – in the same manner that the Busy Beaver function $BB'\left( N \right)$ is in relation to any program. Also, by the very definition of $B{B^ + }_{P{'_T}}$, there cannot be any program of size $ \leqslant N$ running  on ${U_{P{{\text{'}}_T}}}$ that will generate an output higher than or equal to $B{B^ + }_{P{'_T}}\left( N \right)$ -- which may be called the “sub-incompressibility” of the function $B{B^ + }_{P{'_T}}$.

To recreate/relativize the classic uncomputability of the Busy Beaver function, essentially, we had do to Turing and Rad\'{o} the same as Skolem did to Cantor: we relativized the uncomputability of function $BB'\left( N \right)$.  It was demonstrated that it depends on ``being seen from the outside, or from the inside''. Thus, it was proven that there is a ``paradox'' of computability \textit{a la} Skolem, i.e. there is a function that is computable if ``seen from without'', that is uncomputable if ``seen from within''. As both language $L$ and the submachines can be programmed, this phenomenon can occur within our everyday computers.

However, not ``all uncomputabilities'' of a first-order hypercomputer were relativized in relation to a universal Turing machine. Only what was described above was relativized. However, following this line of mathematical inquiry enabled us to build metabiological evolutionary models that are fully analogous to Chaitin’s models of Intelligent Design and Cumulative Evolution -- as shown in [4]. For this purpose, a Turing submachine ${U_{{P^{**}}_T \circ {P_T}}}$ was built and a relative and computable Chaitin Omega number ${\Omega _{{P^{**}}_T \circ {P_T}}}$ – in the case, a time-limited halting probability – was defined.  Clauses were added to ${U_{{P^{**}}_T \circ {P_T}}}$  to allow the existence of finite lower approximations $\rho $ to ${\Omega _{{P^{**}}_T \circ {P_T}}}$ that can be used by a program $P$ when running on ${U_{{P^{**}}_T \circ {P_T}}}$ to compute values of $B{B^ + }_{{P^{**}}_T \circ {P_T}}\left( N \right)$, so that  $2N + C \geqslant \left| {P \circ \rho } \right| \geqslant N + 1$, where $C$ is a constant. This was also designed to mimic what a universal Turing machine can do with lower approximations to $\Omega $ with the purpose of calculating values of $BB'\left( N \right)$. Also, another clauses were added to enable the relative versions of key mutations/programs from Chaitin’s models to also become subprograms. Thus, the open-ended evolution of subprograms revealed itself as isomorphically fast as the open-ended evolution of programs. It allowed us to recursively relativize more ``behaviors'' of a first order hypercomputer in relation to a computer, making them happen between machines and submachines.

An upcoming mathematical inquiry this article suggests is proving whether or not there is a way to define –- relatively -– universal Turing submachines. A universal submachine should be analogous to a universal machine, so there is a class of subcomputable problems, always reducible by subprograms (in the case to the above, within a subcomputable time), that are computable by this universal submachine. The questions would be: how to define a computation time ${P_U}$ so that ${U_{{P_U}}}$ is a universal Turing submachine? Is it possible? This mathematical problem also involves studying the greatest amount of  first order uncomputable functions that can be relativized to become sub-uncomputable. If this Turing submachine is possible, a negative Turing degree can be defined. Moreover, as for each Turing submachine ${U_{P{{\text{'}}_T}}}$ there is always another more powerful and non-reducible submachine ${U_{P{\text{'}}{{\text{'}}_T}}}$ such that  $P{\text{'}}{{\text{'}}_T}$ is sufficient computation time to compute $B{B^ + }_{P{{\text{'}}_T}}$, so it would likewise be possible to create an infinite hierarchy of negative Turing degrees.

To what extent can a computer be made to ``behave'', in relation to one of its subcomputers, as if it was a hypercomputer?

\end{document}